\documentstyle[aps,pra,multicol,epsf]{revtex}
\begin{document}
\title{Natural Thermal and Magnetic Entanglement in 1D Heisenberg Model}
\author{M. C. Arnesen$^1$, S. Bose$^1$ and V. Vedral$^{1,2}$}
\address{$^1$Centre for Quantum Computation, Clarendon Laboratory,
    University of Oxford,
    Parks Road,
    Oxford OX1 3PU, England \\
    $^2$ Optics Section, Blackett Laboratory, Imperial College, Prince Consort Road, London SW7 2BZ, England}
\maketitle

\begin{abstract}
We investigate the entanglement between any two spins in a one
dimensional Heisenberg chain as a function of temperature and the
external magnetic field. We find that the entanglement in an
antiferromagnetic chain can be increased by increasing the
temperature or the external field. Increasing the field can also
create entanglement between otherwise disentangled spins. This
entanglement can be confirmed by testing Bell's inequalities
involving any two spins in the solid.
\end{abstract}


\begin{multicols}{2}

 It is well known that distinct quantum systems can be correlated
 in a "stronger than classical" manner \cite{epr,scr,bell}. This "excess correlation", called entanglement, has
recently become an important resource in quantum information
processing \cite{qcom}. Like energy, it is quantifiable
\cite{bdsw,ved1,ved2}. This motivates us to ask how much
entanglement exists in a realistic system such as a solid (the
likely final arena for quantum computing \cite{kane}) at a finite
temperature. The 1D Heisenberg model \cite{hei1,hei2} is a simple
but realistic \cite{real} and extensively studied
\cite{bethe,bon,yang,egg} solid state system. We analyze the
dependence of entanglement in this system on temperature and
external field. We find that the entanglement between two spins
in an antiferromagnetic solid can be increased by increasing the
temperature or the external field. Increasing the field to a
certain value can also create entanglement between otherwise
disentangled spins. We show that the presence entanglement can be
confirmed by observing the violation of Bell's inequalities.
However, on exceeding a critical value of the field, the
entanglement vanishes at zero temperature and decays off at a
finite temperature. In the ferromagnetic solid, on the other hand,
entanglement is always absent. We compare the entanglement in
these systems to the total correlations.

   The entanglement of formation \cite{bdsw} is
a computable entanglement measure for two spin-$\frac{1}{2}$
systems (qubits) \cite{wot}. We will use this measure to compute
the entanglement between different spins in the 1D isotropic
spin-$\frac{1}{2}$ Heisenberg model. This model describes a
system of an arbitrary number of linearly arranged spins, each
interacting only with its nearest neighbors. Recently,
entanglement in linear arrays of qubits have attracted interest
\cite{wot2,wot3,brig} and in Ref.\cite{wot3} the entanglement in
the ground state of a Heisenberg antiferromagnet has been
computed. But entanglement in the {\em natural} state of a system
as a function of its temperature remains to be studied and the
possibilities of increasing this entanglement by an external
magnetic field remains to be explored. The Hamiltonian for the 1D
Heisenberg chain in a constant external magnetic field $B$, is
given by

\begin{equation}
{\bf H}=\sum_{i=1}^N (B \sigma_z^i + J \vec{\sigma}^i.\vec{\sigma}^{i+1})
\end{equation}
where $\vec{\sigma}^i=(\sigma_x^i, \sigma_y^i, \sigma_z^i)$ in
which $\sigma_{x/y/z}^i$ are the Pauli matrices for the $i$th
spin (we assume cyclic boundary conditions $1+N=1$). $J>0$ and
$J<0$ correspond to the antiferromagnetic and the ferromagnetic
cases respectively. The state of the above system at thermal
equilibrium (temperature $T$) is $\rho(T)=e^{-{\bf H}/kT}/Z$
where $Z$ is the partition function and $k$ is Boltzmann's
constant. To find the entanglement between any two qubits in the
chain, the reduced density matrix $\rho^r(T)$ of those two qubits
is obtained by tracing out the state of the other qubits from
$\rho(T)$. Entanglement is then computed from $\rho^r(T)$
following Ref.\cite{wot}. As $\rho(T)$ is a thermal state, we
refer to this kind of entanglement as {\em thermal entanglement}.
Thermal entanglement is expected to behave differently from the
usual solid state quantities (magnetization, correlations etc.),
as the entanglement of a mixture of states is often less than and
at most equal to the average of the entanglement of these states.



\begin{figure}
\begin{center}
\leavevmode \epsfxsize=8cm \epsfbox{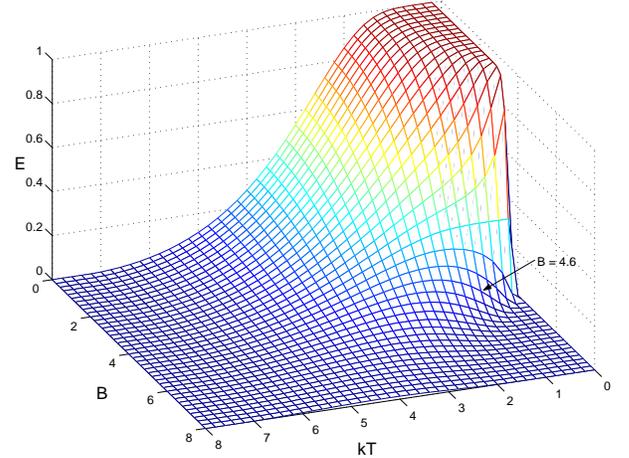}
\caption{
 We have plotted the entanglement $E$ between two qubits
 interacting according to the antiferromagnetic Heisenberg model as a function
of the external field $B$ and temperature (multiplied by the
Boltzmann's constant) $kT$ with coupling $J=1$. The $B=4.6$ line
pointed out in the figure shows that for certain values of $B$ it
is possible to increase $E$ by increasing $T$. At $T=0$,  $E$ has
a sharp transition from $1$ to $0$ as $B$ crosses the critical
value of $B_c=4$. $E$ always becomes zero for values of $T$
exceeding $T_c= 8/k\ln 3$} \label{two}
\end{center}
\end{figure}

  We first examine the $2$ qubit
antiferromagnetic chain. We will use the entanglement of
formation \cite{bdsw,wot,wot1} to calculate the entanglement of
the two qubits. To calculate this entanglement measure, starting
from the density matrix $\rho$, we first need to define the
product matrix $R$ of the density matrix and its time-reversed
matrix
\begin{equation}
  \label{eq:productmatrix}
  R\equiv\rho(\sigma_y\otimes\sigma_y)\rho^*(\sigma_y\otimes\sigma_y).
\end{equation}
Now concurrence is defined by
\begin{equation}
  \label{eq:concurrence}
  C = max\{\lambda_1-\lambda_2-\lambda_3-\lambda_4,0\}
\end{equation}
where the $\lambda_i$ are the square roots of the eigenvalues of
$R$, in decreasing order. In this method the standard basis,
$\{|00\rangle,|01\rangle,|10\rangle,|11\rangle\}$ must be used.
The entanglement of formation is a strictly increasing function
of concurrence, thus there is a one-to-one correspondence. The
amount of entanglement in our special case is given by
\begin{eqnarray}
E&=&-(\frac{1+\sqrt{1-C^2}}{2}) \log_2{(\frac{1+\sqrt{1-C^2}}{2})} \nonumber \\
&-& (\frac{1-\sqrt{1-C^2}}{2}) \log_2{(\frac{1-\sqrt{1-C^2}}{2})},
\end{eqnarray}
where $C$ is the concurrence given by
\begin{eqnarray}
C&=&0~\text{if}~e^{8J/kT}\leq 3, \nonumber \\
&=&\frac{e^{8J/kT}-3}{1+e^{-2B/kT}+e^{2B/kT}+e^{8J/kT}}
~\text{if}~ e^{8J/kT}>3.
\end{eqnarray}
Fig.\ref{two} shows the plot of this entanglement as a function
of magnetic field and temperature.

\begin{figure}
\begin{center}
\leavevmode \epsfxsize=8cm \epsfbox{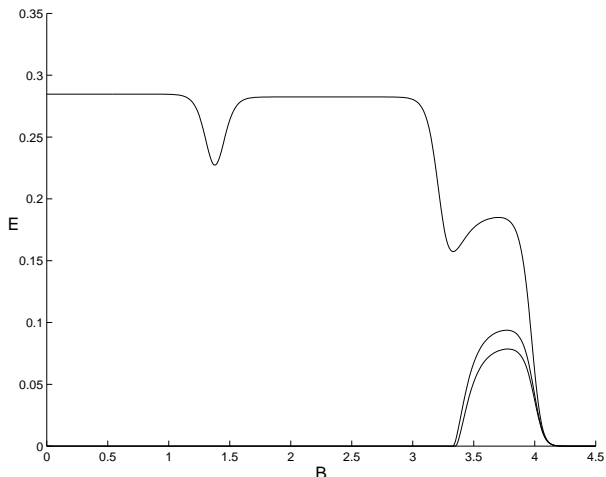}
\caption{
The topmost plot shows the variation of nearest neighbor
entanglement $E$ with $B$ for $N=6$, $kT=0.1$ and $J=1$. The
middle and the bottom most plot show the same for next nearest
and next to next nearest neighbors respectively. The reason for
the shapes of the curves is presented in the text. Note that $l_E$
is $1$ lattice spacing for all values of $B$ below $B_E=3.24$ and
changes to $3$ lattice spacings for a range of $B$ after $B_E$.
This means one can magnetically tune in the entanglement between
any two qubits
 by increasing $B$. We also see the decay of all types of entanglement shortly after $B=B_c=4$.}
\label{elength}
\end{center}
\end{figure}

\begin{figure}
\begin{center}
\leavevmode \epsfxsize=8cm \epsfbox{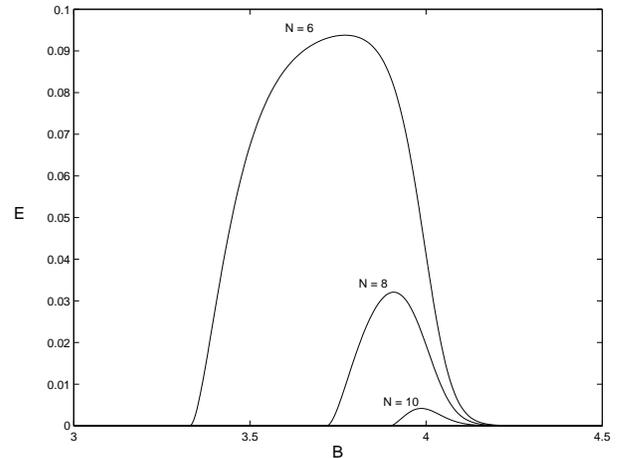}
\caption{We have plotted the variation of next to nearest
neighbor entanglement $E$ with $B$ at $kT=0.1$ and $J=1$ for
three values of $N$. We see that greater the $N$, a larger value
of $B$ is needed to tune in this entanglement, which is absent
until $B$ reaches a certain value. However, this entanglement
disappears irrespective of $N$ shortly after B exceeds $B_c=4$.}
\vspace*{-1cm}
 \label{nneigh}
\end{center}
\end{figure}
For $B=0$, the singlet is the ground state and the triplets are
the degenerate excited states. In this case, the maximum
entanglement is at $T=0$ and it decreases with $T$ due to mixing
of the triplets with the singlet. For a higher value of $B$,
however, the triplet states split, and $|00\rangle$ becomes the
ground state. In that case there is no entanglement at $T=0$, but
increasing $T$ increases entanglement by bringing in some singlet
component into the mixture. On the other hand, as $B$ is
increased at $T=0$, the entanglement vanishes suddenly as $B$
crosses a critical value of $B_c=4J$ when $|00\rangle$ becomes the
ground state. This special point $T=0, B=B_c$, at which
entanglement undergoes a sudden change with variation of $B$, is
the point of a {\em quantum phase transition} \cite{sach}(phase
transitions taking place at zero temperature due to variation of
interaction terms in the Hamiltonian of a system). At any finite
$T$, however, entanglement decays off analytically after $B$
crosses $B_c$. In the ferromagnetic case, the state of the system
at $B=0$ and $T=0$ is an equal mixture of the three triplet
states. This state is disentangled \cite{ved2}. Increasing $B$
increases the proportion of $|00\rangle$ in the state which
cannot make it entangled. Increasing $T$ increases the proportion
of singlet in the state which can only decrease entanglement by
mixing with the triplet. Thus we never find any entanglement in
the $2-$qubit ferromagnet. These features of the $2-$qubit
Heisenberg model are also present in the $N$ qubit model (which we
 investigate numerically) along with additional features,
which we describe next.

     We first plot (Fig.\ref{elength}) how the entanglement between nearest, next nearest
and next to next nearest neighbors in an antiferromagnet vary with
$B$ for a finite but low $T$ (so that the entanglement is
predominantly determined by the ground state). For the nearest
neighbor entanglement there are dips in the entanglement at
certain points. These dips are due to the mixing of two different
entangled ground states at these points. After exceeding a
certain value of $B$ (say, $B_{E}$, which might depend on $N$),
an equal superposition of states with only one spin up becomes
the ground state. This state
$|\Psi_{\mbox{sym}}\rangle=\frac{1}{\sqrt{N}}(|100...0\rangle+|010...0\rangle+...+
|000...1\rangle)$ has entanglement between any two pairs. Thus we
see the next nearest and the next to next nearest neighbor
entanglement becoming finite only after $B$ crosses $B_E$. One
can call this entanglement between non nearest neighbors {\em
magnetic entanglement} as it is brought about  by increasing $B$.
 When $B$ is increased further, beyond a critical value
$B_c=2 J\{1+\frac{1+(-1)^N}{2}+\frac{1-(-1)^N}{2}\cos{\pi/N}\}
\leq 4J$ the disentangled state $|00...0\rangle$ becomes the
ground state. At precisely $T=0$, crossing $B_c$ ensures the
complete vanishing of all types of entanglement. For finite $T$,
all types of entanglement decay to zero gradually after $B_c$.
This is illustrated in Fig.\ref{nneigh}. An interesting point,
shown by all our numerical evidence, is that the change in
entanglement $\Delta E$ at constant temperature due to change in
$B$, can never exceed $|\Delta B|/kT$. This might not be
surprising as entanglement is an entropic quantity and $|\Delta
B|$ is the change in internal energy.

\begin{figure}
\begin{center}
\leavevmode \epsfxsize=8cm \epsfbox{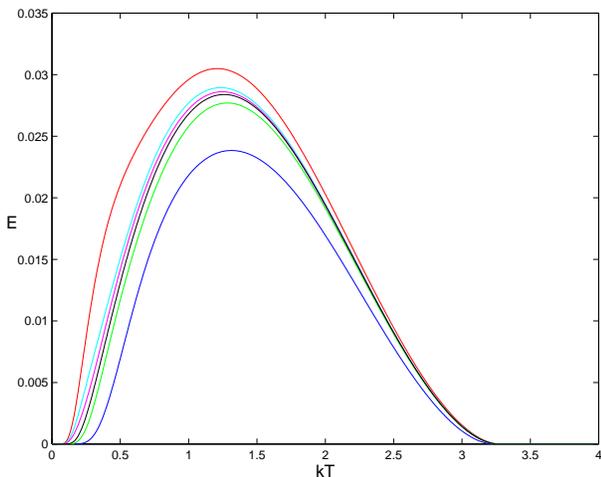}
\caption{
The nearest neighbor entanglement $E$ is plotted as a function of
$kT$ at $B=4.2$, $J=1$ and various values of $N$ (from top to
bottom, $N=6,N=8,N=10,N=9,N=7$ and $N=5$). This graph shows that
$E$ of even $N$ states decreases, $E$ of odd $N$ states increases
with $N$ and they both tend to merge with each other for high $N$
(they almost coincide for $N=9$ and $N=10$). The plot also
illustrates that one can increase the entanglement by increasing
$T$.} \label{temp}
\end{center}
\end{figure}

   If we define a quantity called the {\em entanglement length} $l_{E}$
as the smallest separation between qubits beyond which the
entanglement disappears, then for a small range of $B$ after
crossing $B_E$, $l_{E}$ becomes equal to the farthest neighbor
separation (i.e it can be made arbitrarily large). We have
checked this numerically upto $N=13$, and it is reasonable to
conjecture that this will be true for any $N$. If this conjecture
is false, it will still be interesting to find the value of $N$
beyond which you can never increase $l_E$ to the largest neighbor
separation. Of course, as evident from Fig.\ref{elength}, the
further the qubits are, lesser is the magnitude of the
entanglement between them. Note that the above definition of
entanglement length differs from that defined in Ref.\cite{dorit}
where quantum to classical transitions in noisy quantum computers
was studied (see also Ref.\cite{paivi} for transitions in quantum
networks).

\begin{figure}
\begin{center}
\leavevmode \epsfxsize=8cm \epsfbox{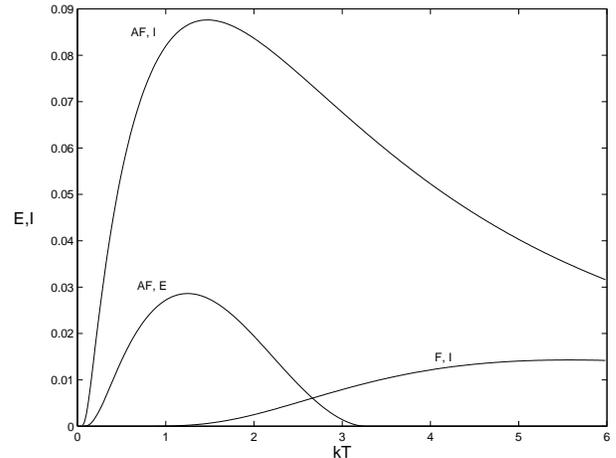}
\caption{
This graph shows the variation of total mutual information with
temperature for the antiferromagnetic (AF,I) and ferromagnetic
(F,I) case for $N=10$, $B=4.2$ and $|J|=1$. The entanglement for
the antiferromagnetic case (AF, E) is also plotted for a
comparison. } \label{mutin}
\end{center}
\end{figure}

   As mentioned earlier, $|\Psi_{\mbox{sym}}\rangle$ becomes
the ground state for a certain range of values of the external
magnetic field (confirmed numerically up to $N=13$ and conjectured
for other values of $N$). At extremely low temperatures and
appropriate magnetic fields, thus, the state of the chain will
almost be $|\Psi_{\mbox{sym}}\rangle$. This state has the
interesting property that there exists entanglement between any
two qubits. The reduced density matrix of any two spins in the
state $|\Psi_{\mbox{sym}}\rangle$ is
$\rho=\frac{2}{N}|\Psi^{+}\rangle\langle\Psi^{+}|+(1-\frac{2}{N})|00\rangle
\langle 00|$ where
$|\Psi^{+}\rangle=\frac{1}{\sqrt{2}}(|01\rangle+|10\rangle)$. This
is an entangled state. For any $N$, if we measure the state of all
qubits except two, those two qubits would be projected onto a
maximally entangled state, which can then be verified through
Bell-CHSH inequalities \cite{horo}. Even for any other mixed state
which may thermally or magnetically generated, there exists A
neccessary and sufficient condition to check whether the CHSH
inequality is violated \cite{horo}. Of course, one has to make an
appropriate choice of measurement axes on the two spins in the
solid. As different components of the magnetic susceptibility
tensor are proportional to spin-spin correlations in different
pairs of directions \cite{sach}, a CHSH inequality can tested by
measuring different components of the magnetic susceptibility
tensor .


We now look at the dependence of entanglement on $T$ in the $N$
qubit case for a fixed $B$. Fig.\ref{temp} shows that one can
increase entanglement by increasing $T$.  After a certain $T$,
all entanglement dies out. In all simulations we find this
temperature to be lower than $T_c=8/\ln3$. Also, we see that the
curves for entanglement in the case of even and odd $N$ approach
each other as $N$ increases. This seems reasonable because for
large $N$, it should not make a difference to the nearest
neighbor entanglement whether we add or subtract a qubit
somewhere far in the chain. As with the 2-qubit case, we find no
thermal or magnetic entanglement in a ferromagnetic chain.



   We would now like to compare the amount of entanglement in the solid
to the {\em total} two qubit correlations. An information
theoretic measure of these correlations is the mutual information
given by $I(i:j)=S(\rho^i)+S(\rho^j)-S(\rho^{ij})$ where
$\rho^i$, $\rho^j$  are the density matrices of the $i$th and
the  $j$th spin respectively, $\rho^{ij}$ is their joint state
and $S(\rho)$ represents the entropy of $\rho$. In a manner
similar to  the {\em connected} correlation function
\cite{binney}, this quantity measures the effect that genuinely
results from the interaction between particles. A plot of
$I(i:j)$ with temperature is shown in Fig.\ref{mutin}. It is
interesting to note that though entanglement is always absent in
a ferromagnet, $I$ for nearest neighbors (stemming entirely from
classical correlations) can be increased by increasing the
temperature. It is well known that the magnetic susceptibility is
proportional to the spin-spin correlations \cite{egg}. It would be
interesting to investigate whether any difference arises between
the antiferromagnetic and ferromagnetic susceptibility tensors
due to the complete absence of entanglement in the latter case.

In this letter, we have introduced the concepts of thermal and
magnetic  entanglement and analyzed their behaviour in the $1$D
isotropic Heisenberg model. We have found critical values of
field beyond which entanglement disappears at zero temperature
and declines at finite temperature. Our results indicate that
there is also a critical temperature after which all entanglement
vanishes, though there is a range of field in which entanglement
can be increased by increasing the temperature. Based on numerical
evidence, we have conjectured that the entanglement length can be
made arbitrarily large by applying an appropriate external
magnetic field. We have also compared the total correlations to
the entanglement. Our work raises a number of interesting
questions and conjectures to prove and the possibility of
numerous generalizations such as higher dimensions, non nearest
neighbor interactions, anisotropies, other Hamiltonians and so on.
In addition, we also showed that by applying a suitable magnetic
field and lowering the temperature sufficiently, and doing
suitable projections, one can create a state which violates the
Bell's inequalities. The "natural" entanglement can be verified
in these cases. In future we will investigate how to map this
natural entanglement out of the solid (eg. by neutron scattering)
and use it as a resource in communications.

\vspace*{.2cm}

  We thank M. Bourenanne and V. Kendon for useful
discussions. After completion of the article M. Nielsen brought
our attention to the fact that the two qubit case has been
considered previously by him in \cite{Nielsen}. Funding by the
European Union project EQUIP (contract IST-1999-11053) and
   Hewlett Packard is acknowledged. MCA acknowledges funding from
   George F. Smith through
the Associates SURF endowment.
   VV acknowledges hospitality of the Erwin Schr\"{o}dinger
   Institute in Vienna where part of the work was carried out.

\end{multicols}
\end{document}